\documentclass[runningheads]{llncs}
\usepackage{amsmath}
\usepackage{amssymb}
\usepackage{marvosym} 

\usepackage{multirow}
\usepackage{utfsym}

\usepackage[T1]{fontenc}
\usepackage[colorlinks,linkcolor=blue,citecolor=blue,urlcolor=blue]{hyperref}

\usepackage{graphicx,verbatim}

\begin{document}

\title{APSeg: Auto-Prompt Model with Acquired and Injected Knowledge for Nuclear Instance Segmentation and Classification}

\author{Liying Xu\inst{1} \and
Hongliang He\inst{1}\textsuperscript{({\Letter})}  \and
Wei Han\inst{1} \and
Hanbin Huang\inst{1} \and
Siwei Feng\inst{1} \and
Guohong Fu\inst{1} 
}
\authorrunning{Xu et al.}
\titlerunning{APSeg}
%
\institute{School of Computer Science and Technology, Soochow University \\
\email{hlhe2023@suda.edu.cn}
}
    
\maketitle 
\begin{abstract}
Nuclear instance segmentation and classification provide critical quantitative foundations for digital pathology diagnosis. With the advent of the foundational Segment Anything Model (SAM), the accuracy and efficiency of nuclear segmentation have improved significantly. However, SAM imposes a strong reliance on precise prompts, and its class-agnostic design renders its classification results entirely dependent on the provided prompts. Therefore, we focus on generating prompts with more accurate localization and classification and propose \textbf{APSeg}, \textbf{A}uto-\textbf{P}rompt model with acquired and injected knowledge for nuclear instance \textbf{Seg}mentation and classification. APSeg incorporates two knowledge-aware modules: (1)  Distribution-Guided Proposal Offset Module (\textbf{DG-POM}), which learns distribution knowledge through density map guided, and (2) Category Knowledge Semantic Injection Module (\textbf{CK-SIM}), which injects morphological knowledge derived from category descriptions. We conducted extensive experiments on the PanNuke and CoNSeP datasets, demonstrating the effectiveness of our approach. The code will be released upon acceptance. 

\end{abstract}

\section{Introduction}

Accurate and automated nuclear segmentation and classification is fundamental for clinical pathological image analysis and diagnosis, providing important quantitative results for tasks such as survival rate prediction and cancer staging and grading \cite{greenwald2022whole,abdel2023comprehensive}. Segmentation of densely packed, small-volume nuclei and the recognition of subtle category differences is a non-trivial task. Traditional deep learning methods often rely on encoder-decoder structures like U-Net \cite{ronneberger2015u}, which design different task-aware branches based on the results, such as auxiliary vertical and horizontal distance maps \cite{graham2019hover}, distance maps \cite{he2021cdnet}, and topological information \cite{he2023toposeg}, to achieve bottom-up instance segmentation and classification with multiple map-assisted methods. However, these methods rely on carefully crafted post-processing, requiring detailed tuning of hyperparameters \cite{yao2023pointnu}.

Recently, owing to the generalization capabilities of the large vision model SAM \cite{kirillov2023segment}, fine-tuning based on pre-trained models has greatly reduced the dependency on large datasets. Many SAM-based medical image segmentation models have achieved state-of-the-art (SOTA) performance in areas such as polyps \cite{li2024asps,xiong2024sam2}, tumors \cite{qin2024db,shi2024mask,xiong2024sam2}, lesions \cite{gu2024lesam,wu2023medical,li2024tp}, organs \cite{cheng2024unleashing}, and surgical instruments \cite{yue2024surgicalsam}. In the field of pathological images, CellViT \cite{horst2024cellvit} uses pre-trained encoded features as additional knowledge injection, still applying the traditional U-Net architecture; SAC \cite{na2024segment} implements an automatic prompt semantic segmentation framework, but still depends on expert prompts and does not achieve instance segmentation; SAMAug \cite{zhang2023input} leverages SAM's zero-shot capability for image augmentation, significantly improving the instance segmentation performance of expert models; PromptNucSeg \cite{shui2025unleashing} leverages the interactive segmentation capabilities of SAM, achieves a prompt-driven instance segmentation model. However, SAM-based instance segmentation introduces a severe dependency on precise prompts.

To address this issue, we focus on designing a nuclear auto-prompter that better adapts to SAM by learning representations of the pathological image microenvironment. Thus, we propose APSeg: an auto-prompt model with acquired and injected knowledge for nuclear instance segmentation and classification. In APSeg, we introduce two knowledge-aware modules: Distribution-Guided Proposal Offset Module (DG-POM) leverages the nuclear counting task to help the model acquire knowledge of nuclear distribution in pathological images, generating distribution-guided deformed proposals to enhance the sampling features for regression and classification tasks; and Category Knowledge Semantic Injection Module (CK-SIM) focuses on injecting detailed nuclear morphological knowledge using the pre-trained CLIP model \cite{radford2021learning}, thereby improving the classification accuracy of prompts.

In summary, our contributions are as follows: (1) We build an auto-prompt model named APSeg, a novel nuclear auto-prompter designed for nuclear instance segmentation and classification. (2) We propose two modules, DG-POM and CK-SIM,  separately augment nuclear detection and classification precision through acquired and injected knowledge. (3) We conduct extensive experiments on the PanNuke and CoNSeP datasets, achieving state-of-the-art performance.

\section{Method}
\subsubsection{Problem Definition.} Given an input image $ I \in \mathbb{R}^{H \times W \times 3}$ and the corresponding prompts $P = \{p_i\}_{i=1}^{N} $ where $p_i = ((x_i,y_i),c_i)$, and $c_i \in C$, labeling the position $(x_i,y_i)$ and category $c_i$ of the $i$-th instance. Nuclear instance segmentation is required to follow the prompts to identify each individual instance $i$ and produce the result map $ M_i \in \mathbb{R}^{H \times W}$ , where each pixel \( M_i(x,y) \in \{0,1,2,\dots, i,\dots,N\} \) represents the instance identifier to which the pixel belongs, with \( 0 \) denoting the background and \( i \) representing a specific instance ID. Our core focus is on how to obtain prompts $P$ that are more accurate in both localization and classification to guide nuclear segmentation.

\subsubsection{Framework Overview.} 
\begin{figure}[t]
\includegraphics[width=\textwidth]{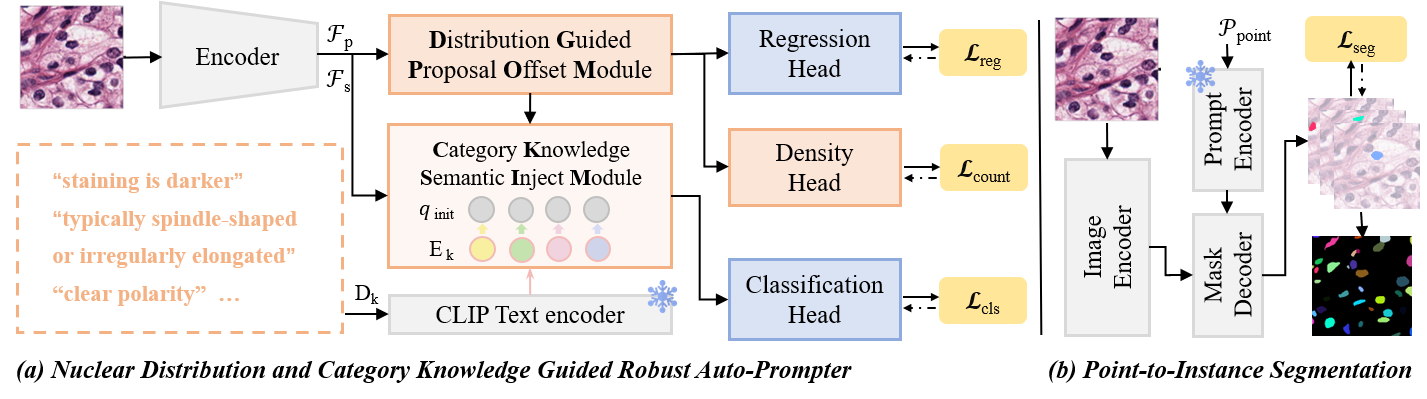}
\caption{Overview of APSeg. (a) APSeg consists of two knowledge-aware modules DG-POM and CK-SIM, with a (b) point-to-instance SAM for instance segmentation followed the prompt generated by APSeg.} \label{fig1}
\end{figure}

The novel auto-prompt framework APSeg is as figured in Fig.\ref{fig1}(a), the Segment Anything Model used for nuclear segmentation followed its original design as figured in Fig.\ref{fig1}(b).  We adopt the method from \cite{shui2025unleashing} to design a fine-tuned SAM for point-to-instance task modeling, where all point prompts during testing are generated by APSeg. APSeg introduces an acquired and injected knowledge guided approach through two functional modules DG-POM and CK-SIM.

\subsection{Distribution Guided Proposal Offset Module} 

Different nuclei types across tissues exhibit diverse distribution patterns and densities. Under such complex distributions, fixed proposals make the regression task challenging. DG-POM is designed to overcome this challenge. 

\begin{figure}[t]
\includegraphics[width=\textwidth]{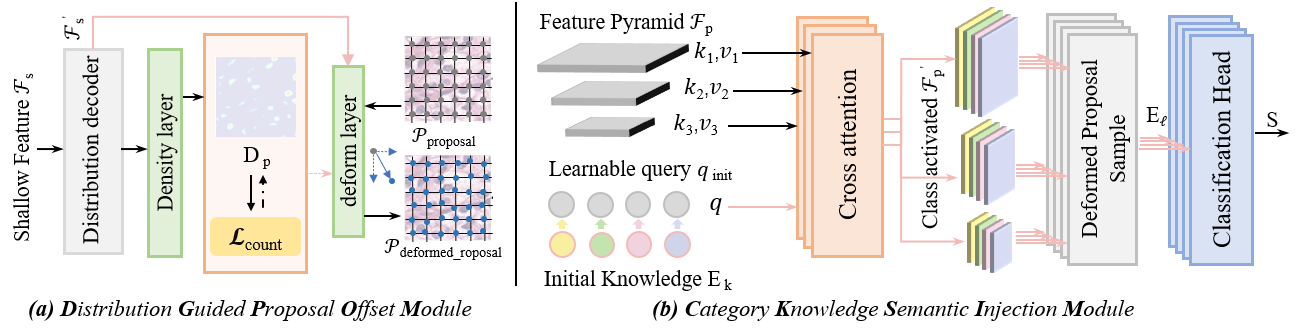}
\caption{ The detailed architectures of (a) Distribution Guided Proposal Offset module (DG-POM) and (b) Category Knowledge Semantic Injection Module (CK-SIM). The parts of knowledge acquisition and injection are highlighted with \textcolor{pink}{pink} lines.} \label{fig2}
\end{figure}

As illustrated in Fig.\ref{fig2}(a), inspired by studies in counting tasks \cite{sun2023indiscernible,kang2024vlcounter}, density maps effectively capture instance distribution. Therefore, we introduce an additional density map based nuclear counting task in APSeg. The original proposal design is modified as follows: 
\begin{align}
 F_{p} &= \phi_{backbone}(I)\\
    F_{s}’ &= \phi_{relu}(\phi_{conv}(\phi_{relu}(\phi_{conv}(F_{s})))) \\
     (\Delta X_p,\Delta Y_p) &= \phi_{deform\_layer}(sample(F_{s}',P_{proposal})) \\
     P_{deformed\_proposal} &= P_{proposal} +   (\Delta X_p,\Delta Y_p)
\end{align}

Shallow feature $F_s$ which extracted from feature pyramid $F_p$ contains more positional information. It is processed by a distribution decoder, composed of stacked convolutional and activation layers, to obtain an updated feature map $F_s'$. This updated $F_s'$ is used for sampling the initial proposals $P_{proposal}$ and then passed through a separate deform layer $\phi_{deform\_layer}(\cdot)$  to compute the proposal offsets $ (\Delta X_p,\Delta Y_p) $. 

The deformed proposals capture the initial nuclear distribution bias, alleviating the burden of regressing real nuclei locations.
\begin{align}
    E_{p} &= sample(F_{p},P_{deformed\_proposal}) \\
     (\Delta X,\Delta Y) &= \phi_{reg\_head}( E_{p}) \\
     P_{point} &= P_{deformed\_proposal} +   (\Delta X,\Delta Y)
\end{align}

The predicted points $P_{point}$ matched to ground truth points $P_{GT}$ are supervised, with the matching performed using the Hungarian algorithm.

To enable this process to acquire distribution characteristics, we additionally introduce a density layer $\phi_{density\_layer}(\cdot)$  for generating a density map $D_p \in \mathbb{R}^{1 \times 1 \times \frac{H}{2^2} \times \frac{W}{2^2}}$ nuclear instance counting and incorporate a quantity-based loss for supervision: 
\begin{align}
   D_p &= \phi_{density\_layer}(F_s') \\
    \mathcal{L} _{count} &= MAE(\sum D_p,N) 
\end{align}

The supervision from the density map regression task and nuclear counting enables the deform layer to effectively learn instance distribution knowledge.

\subsection{Category Knowledge Semantic Injection Module }
Pathology images contain complex semantic microenvironment information, where nuclei of different categories exhibit subtle morphological differences that are difficult to distinguish, leading to inter-class homogeneity. Inspired by the multimodal pre-trained model CLIP \cite{radford2021learning} aligns visual and textual semantics, we leverage text to explicitly inject general feature semantic knowledge, aiding in distinguishing inter-class differences and improving nucleus classification.

The pre-trained model CLIP, with general visual-text semantic capabilities, excels at representing basic semantics. Instead of using specialized terminology for nucleus categories, we choose more intuitive morphological descriptions to construct morphological and microenvironmental knowledge for different categories. This knowledge is represented as $E_{k} = \phi_{text\_encoder}(D_{k}) \in \mathbb{R}^{C \times C_{k}}$, where  $D_k$ is the description knowledge of $C$ categories, and $C_k$ is the dimensionality of the knowledge text features.

As illustrated in Fig.\ref{fig2}(b), we design a method that utilizes a learnable class-aware query $q \in \mathbb{R}^{C \times C_k}$ to activate the feature pyramid $F_p$ and apply $E_k$ as prior knowledge to initialize $q$. This enables a class-aware query-based comprehensive classification evaluation, ultimately optimizing classification performance. The process is as follows:
\begin{align}
   q' = w_qq  ,\; k^l &= w_kF^l_p ,\;v^l = w_vF^l_p ,\; l \in \{1,2,3\}  \\
   A^l &= Softmax(\frac{q'(k^l)^T}{\sqrt{d}}) \\
   (F^l_p)' &=A^l v^l
\end{align}

where $k$ and $v$ come from the multi layer of the feature pyramid $F_p$. 

After obtaining the class-activated $F_p' \in \mathbb{R}^{ C\times C_I \times \frac{H}{2^{i}} \times \frac{W}{2^{i}}} , i\in\{2,3,4\}$ using the class-aware query, the classification is determined separately for the multi-scale feature maps of $C$ classes, and the results are aggregated:
\begin{align}
   E_l^i&= Sample(F_l^i,P_{deformed\_proposal}^i) ,\;i \in\{1,2,..., C\}\\
   S &= \phi_{cls\_head}(\phi_{conv}(Concat(E^i_l)))
\end{align}
where classification result $S \in \mathbb{R}^{N \times (C+1)}$ is supervised using the cross-entropy loss, where $w_c$ is the weight of each categories and $c_n$ is the truth category:
\begin{equation}
       \mathcal{L}_{cls} = - \sum^N_{n=1} w_{c}log(\frac{e^{S_n,c_n}}{\sum_{c=1}^Ce^S_{n,c}})  \\
\end{equation}

Multi-class aggregation classification, achieved through a category-aware query-activated feature pyramid, ensures the injection of category knowledge.

\begin{table}[t]
    \centering
    \caption{Performance comparison on the PanNuke dataset, both binary PQ (bPQ) and multi-class PQ (mPQ) are computed. The best and second-best scores are highlighted in\textbf{ bold} and \underline{underlined}. The results of PNS(PromptNucSeg) are provided by us.}
    \begin{tabular}{c|cc|cc|cc|cc|cc}
    \hline
         & \multicolumn{2}{|c|}{\textbf{HoVer-Net}} & \multicolumn{2}{|c|}{\textbf{PointNu-Net}} & \multicolumn{2}{|c|}{\textbf{CellViT-H}}  & \multicolumn{2}{|c|}{\textbf{PNS-H}} & \multicolumn{2}{|c}{\textbf{APSeg}} \\ \hline
        Tissue & bPQ &  mPQ & bPQ &  mPQ & bPQ &  mPQ & bPQ &  mPQ & bPQ &  mPQ  \\ \hline
        Adrenal & 69.62 & 48.12 & 71.34 & \underline{51.15} & 70.86 & \textbf{51.34} & \underline{71.53}& 50.06 
& \textbf{72.10} &   50.57 
\\ 
        Bile Duct & 66.96 & 47.14 & 68.14 & 48.68 & 67.84 & 48.87 & \underline{68.98} & \underline{49.21} 
& \textbf{69.52} &   \textbf{49.82}
\\ 
        Bladder & 70.31 & 57.92 & \underline{72.26} & \textbf{60.65} & 70.68 & 58.44 & \textbf{72.38} & \underline{59.80} 
& 71.78 &   58.94 
\\ 
        Breast & 64.70 & 49.02 & 67.09 & 51.47 & 67.48& 51.80& \textbf{68.30} & \underline{52.07} 
& \underline{68.11} &   \textbf{52.60}
\\ 
        Cervix & 66.52 & 44.38 & 68.99 & \underline{50.14} & 68.72 & 49.84 & \underline{69.39} & 50.12
& \textbf{69.51} &   \textbf{50.88}
\\ 
        Colon & 55.75 & 40.95 & 59.45 & 45.09 & 59.21 & 44.85 & \underline{60.55} & \underline{45.38} 
& \textbf{60.75} &   \textbf{46.21} 
\\ 
        Esophagus & 64.27 & 50.85 & 67.66 & 55.04 & 66.82 & 54.54 & \underline{68.17} & \underline{55.37} 
& \textbf{68.54} &   \textbf{55.63} 
\\ 
        Head \& Neck & 63.31 & 45.30 & 65.46 & 48.38 & 65.44 & 49.13 & \underline{66.03} & \underline{50.07} 
& \textbf{67.02} &   \textbf{50.25} 
\\ 
        Kidney & 68.36 & 44.24 & 69.12 & 50.66 & 70.92 & 53.66 & \textbf{71.19} & \underline{54.18} 
& \underline{70.94} &   \textbf{56.38} 
\\ 
        Liver & 72.48 & 49.74 & 73.14 & 51.74 & 73.32 & 52.24& \underline{73.45} & \underline{52.33} 
& \textbf{73.64} &   \textbf{53.11}
\\ 
        Lung & 63.02 & 40.04 & 63.52 & 40.48 & 64.26& \underline{43.14} & \underline{65.49} & 43.01 
& \textbf{65.58} &   \textbf{43.61} 
\\ 
        Ovarian & 63.09 & 48.63 & \textbf{68.63} & \textbf{54.84} & 67.22& 53.90& 68.23 & 53.63 
& \underline{68.31} &   \underline{54.21} 
\\ \
        Pancreatic & 64.91 & 46.00 & 67.91 & 48.04 & 66.58& 47.19 & \textbf{68.79} & \underline{49.34} 
& \underline{68.25} &   \textbf{49.61} 
\\ 
        Prostate & 66.15 & 51.01 & 68.54 & 51.27 & 68.21& 53.21 & \textbf{69.68} & \textbf{54.73} 
& \underline{69.34} &   \underline{53.77} 
\\ 
        Skin & 62.34 & 34.29 & 64.94 & 40.11 & 65.65& \textbf{43.39} & \underline{66.71} & \underline{41.52} 
& \textbf{66.75} &   41.38 
\\ 
        Stomach & 68.86 & \textbf{47.26} & 70.10 & 45.17 & 70.22& \underline{47.05}& \textbf{71.38} & 46.32 
& \underline{71.23} &   45.79 
\\ 
        Testis & 68.90 & 47.54 & 70.58& \textbf{53.34}& 69.55& 51.27 & \textbf{70.99} & 51.73 
& \underline{70.51} &   \underline{53.08} 
\\ \
        Thyroid & 69.83 & 43.15 & 70.76& 45.08& \underline{71.51}& \underline{45.19} & 71.28 & 44.50 
& \textbf{72.31} &   \textbf{45.96} 
\\ 
        Uterus & 63.93 & 43.93 & 66.34& \underline{48.46} & 66.25& 47.37& \underline{67.15} & 46.89 
& \textbf{67.20} &   \textbf{48.72} 
\\ \hline
        Average & 65.96 & 46.29 & 68.08 & 49.57 & 67.93 & 49.80& \underline{68.93} & \underline{50.01} & \textbf{69.02} &   \textbf{50.55} \\
        Std & 0.36 & 0.76 & 0.50& 0.82& 3.18 & 4.13 & 0.10 & 0.25 & 0.09 & 0.23 \\ \hline
    \end{tabular}
    \label{table1}
\end{table}

\subsection{Training Loss} 
The training loss function of APSeg is obtained by weighting the losses of nuclear classification, points regression and instance counting tasks: 
\begin{equation}
\mathcal{L} = w_{cls}\mathcal{L}_{cls} + w_{reg}\mathcal{L}_{reg} + w_{count}\mathcal{L}_{count}
\end{equation}
 
 The regression task is supervised using the L1 loss, where $p_n$ is the location of the $n$-th nuclei and $\hat{p_n}$ is its ground truth position: 
\begin{equation}
\mathcal{L}_{reg} = \sum_{n=1}^{N}||p_n - \hat{p_n}||_1
\end{equation}

\section{Experiments}
\subsection{Experimental Setup}
\subsubsection{Dataset.} 
PanNuke is the largest publicly available dataset for nuclei segmentation and classification \cite{gamper2020pannuke,gamper2019pannuke}, sourced from The Cancer Genome Atlas (TCGA) and spanning 19 tissue types. It contains 256×256 patches extracted from over 20,000 Whole Slide Images, divided into three folds for training, validation, and testing. Following \cite{gamper2020pannuke}, we replicate the results using the same methodology. Annotations are semi-automatically generated and quality-checked by clinical pathologists to ensure statistical alignment with real-world clinical scenarios and minimal selection bias. CoNSeP \cite{graham2019hover} consists of 41 images with diverse cell types, categorized into four classes: miscellaneous, inflammatory, epithelial, and spindle-shaped. Following prior studies, it is split into 27 training and 14 testing images.

\begin{table}[!ht]
\centering
    \fontsize{10pt}{10pt}\selectfont 
    \centering
    \caption{Average PQ across three folds for each nuclear category on the PanNuke.}
    \begin{tabular}{c|ccccc|c}
    \hline
        Method & Neoplastic & Epithelial&  Inflammatory& Connective& Dead   &Average\\ \hline
        Mask-RCNN & 0.472 & 0.403 & 0.290 & 0.300 & 0.069  &0.307 
\\ 
        DIST & 0.439 & 0.290 & 0.343 & 0.275 & 0.000  &0.269 
\\
        StarDist & 0.547 & 0.532 & 0.424 & 0.380 & 0.123  &0.401 
\\
        Micro-Net & 0.504 & 0.442 & 0.333 & 0.334 & 0.051  &0.333 
\\
        HoVer-Net & 0.551 & 0.491 & 0.417 & 0.388 & 0.139  &0.397 
\\
        CPP-Net & 0.571 & 0.565 & 0.405 & 0.395 & 0.131  &0.413 
\\
        PointNu-Net & 0.578 & 0.577 & \textbf{0.433} & 0.409 & 0.154  &0.430 
\\
        CellViT-H & 0.581 & \textbf{0.583} & 0.417 & \underline{0.423} & 0.149  &0.431 
\\
        PromptNucSeg-H & \textbf{0.594} & \underline{0.578} & 0.410 & 0.422 & \textbf{0.174}  &\underline{0.436} 
\\  \hline
        APSeg &  \underline{0.592} &\textbf{0.583} & \underline{0.431} & \textbf{0.425} & \underline{0.161}  &\textbf{0.438} \\ \hline
        
    \end{tabular}
    \label{table2}
\end{table}

\begin{table}[h]
    \centering
    \setlength{\tabcolsep}{4pt}
    \caption{Performance comparison on the CoNSeP dataset. The best and second-best scores are highlighted in\textbf{ bold} and \underline{underlined}.}
    \begin{tabular}{c|ccccccccccc}
    \hline
        Method & Dice & AJI & DQ & SQ & PQ & $F_{Det}$ & $F_m$ & $F_i$ & $F_e$ & $F_s$  \\ \hline
        Micro-Net & 79.4 & 52.7 & 60.0 & 74.5 & 44.9  & 74.3 & 11.7 & 59.2 & 61.5 & 53.2  \\ 
        Mask-RCNN & - & - & 61.9 & 74.0 & 46.0  & 69.2 & 9.8 & 59.0 & 59.5 & 52.0  \\ 
        DIST & 80.4 & 50.2 & 54.4 & 72.8 & 39.8  & 71.2 & 0.0 & 53.4 & 61.7 & 50.5  \\ 
        HoVer-Net & \textbf{84.5} & 55.2 & 68.5 & \underline{77.5} & 53.2  & 73.7 & 28.6 & 60.6 & 62.8 & 53.5  \\ 
        PointNu-Net & - & - & \textbf{71.4} & 76.2 & \textbf{55.5}  & 75.2 & 46.2 & 64.7 & 66.1 & 55.9  \\ 
        PromptNucSeg & 81.0 & \underline{54.2} & 68.5 &\textbf{ 77.7 }& 53.3  & 75.8 & 47.7 & \textbf{76.0} & 68.1 & 62.9   \\ \hline
        APSeg & \underline{82.3} &\textbf{55.8} & \underline{70.6} & \underline{77.5}& \underline{54.9} & \textbf{76.8} & \textbf{48.0} & \underline{75.4} & \textbf{69.4} & \textbf{65.4}  \\ \hline
    \end{tabular}
    \label{table3}
\end{table}

\subsubsection{Implementation Details.} Our model is implemented based on the PyTorch framework, and all experiments are performed on a single V100 32GB GPU. For both datasets, we train the APSeg for 300 epochs, with a learning rate set to 1e-4 with a weight decay of 1e-4 and the AdamW optimizer used. We set the class weights as \( w_c = 1 \) for \( c = \{0, \dots, C-1\} \), with the background weight \( w_C = 0.3 \). The loss weights $w_{cls}$, $w_{reg}$ and $w_{count}$ are set to 1.0, 5e-3, and 1e-4, respectively. The encoder uses ConvNeXt \cite{woo2023convnext}. For SAM and CLIP, we employ their ViT-H \cite{dosovitskiy2020image} and RN50x16 variants, respectively. The category knowledge description texts are jointly designed by pathology experts and GPT-4.
\subsection{Comparison Study}

On the PanNuke dataset, we followed the experimental setup of \cite{gamper2020pannuke} and conducted comprehensive evaluations across three folds. We compared multi-class PQ (mPQ) and binary PQ (bPQ) across 19 tissue types, as shown in Tab.\ref{table1}. Our model achieved an overall bPQ of 69.02\% and an mPQ of 50.55\%. Additionally, we evaluated PQ across five specific categories as shown in Tab.\ref{table2}, attaining an average PQ score of 43.8\%, all of which reached state-of-the-art performance.  

On the CoNSeP dataset, we compared five semantic segmentation metrics: Dice, AJI, DQ, SQ, and PQ, along with the F1 scores for four different categories as shown in Tab.\ref{table3}. While our model achieved SOTA classification performance, the point-based detection approach still suffered from significant omissions in the densely packed CoNSeP dataset, leading to some performance gaps in semantic metrics. Nevertheless, our method achieved an improvement of 1.6\% in AJI and PQ compared to the baseline.

\begin{figure}[h]
\centering
\includegraphics[width=\textwidth]{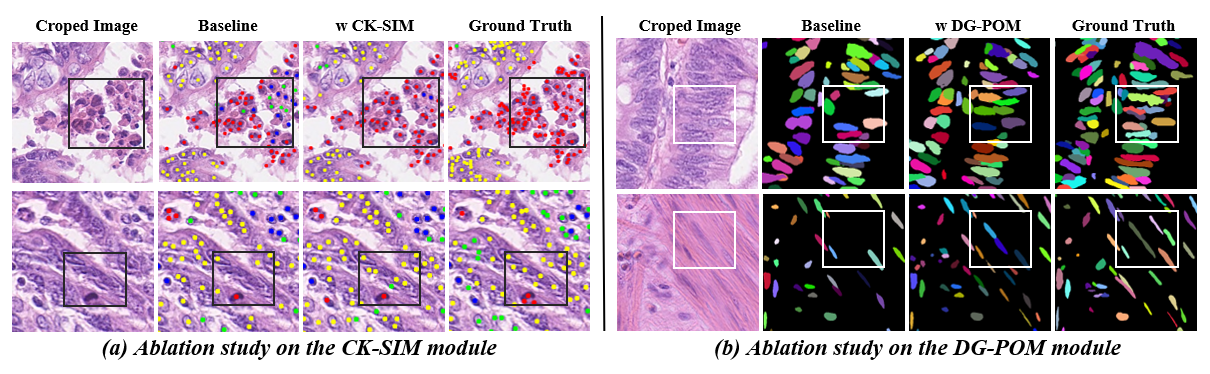}
\caption{The visual comparison of ablation study for CK-SIM(a) and DG-POM(b).} \label{fig3}
\end{figure}

\begin{table}
    \centering
\caption{Ablation study on CoNSeP dataset. The classification and detection results specifically evaluate point prompts. }
\label{tab:my_label}
    \begin{tabular}{cc|ccc|ccc|ccc}
    \hline
\multirow{2}{*}{DG-POM}  & \multirow{2}{*}{CK-SIM} & \multicolumn{3}{c|}{Classification}& \multicolumn{3}{c|}{Detection}& \multicolumn{3}{c}{Segmentation}\\ 
    \cline{3-11}
         & &  P&  R&  F
&  P&  R&  F&   Dice&AJI&PQ
\\ \hline
         &  &  74.68&  61.54&  66.29
&  84.08&  74.08&  78.76&   81.06&54.20&53.38
\\ 
         \usym{2714} &  &  74.95&  \textbf{63.19}&  67.55
&  82.05&  \textbf{77.42}&  79.66&   \underline{82.14}&\underline{55.43}&54.07
\\ 
         &  \usym{2714} &  \textbf{78.48}&  62.83&  \textbf{69.33}
&  87.50&  74.07&  80.22&   80.75&54.29&\underline{54.61}
\\ 
         \usym{2714} &  \usym{2714} &  \underline{75.56}&  \underline{63.01}&  \underline{67.70}&  \textbf{85.07}&  \underline{76.88}&  \textbf{80.77}&   \textbf{82.38}&\textbf{55.83}&\textbf{54.90}\\ \hline
    \end{tabular}
    \label{table4}
\end{table}

\subsection{Ablation Study}
As shown in Tab.\ref{table4}, we comprehensively evaluate the precision (P), recall (R), and F1-score of point detection and classification obtained by APSeg. Due to space limitations, we only present the ablation experiments on CoNSeP. Additionally, we present the instance segmentation results under the same segmenter using Dice, AJI, and PQ metrics. It can be observed that CK-SIM significantly improves point classification, achieving a 3.04\% increase in classification F1-score. Meanwhile, DG-POM enhances instance recall by 3.34\%, which greatly benefits segmentation performance, leading to a 1.23\% improvement in AJI and a 0.69\% increase in PQ. Fig.\ref{fig3} further illustrates the improvements in classification and detection achieved by these two modules. When both modules are used together, the points integrate the advantages of detection and classification, achieving the best segmentation performance.

\section{Conclusion} 
In this paper, we proposed APSeg, an automatic prompt generator for nuclear instance segmentation. APSeg leverages self-acquired knowledge and external knowledge injection to generate more robust prompts. We design and implement a density map-guided proposal offset module and a category knowledge semantic injection module. Extensive comparisons based on the prompt-segment pipeline on the PanNuke and CoNSeP datasets demonstrate the effectiveness and robustness of the prompts generated by APSeg.

\bibliographystyle{splncs04}
\bibliography{reference}
\end{document}